\documentclass[aps,pra,onecolumn]{revtex4-2}

\usepackage{amsmath,amsfonts,qcircuit,bbm,braket}
\usepackage[colorlinks=true,citecolor=red]{hyperref}
\usepackage{graphicx}
\usepackage[caption=false]{subfig}

\newcommand{\spin}[1]{{\bf{S}}_{#1}}
\newcommand{\parameter}[1]{{\boldsymbol{#1}}}
\newcommand{\simulator}[1]{\texttt{#1$\_$simulator}}
\newcommand{\device}[1]{\texttt{ibmq$\_$#1}}

\begin{document}

\title{Quantum circuits for the preparation of spin eigenfunctions on quantum computers}

\author{Alessandro Carbone}
\affiliation{Theory and Simulation of Materials (THEOS), and National Centre for Computational Design and Discovery of Novel Materials (MARVEL), École Polytechnique Fédérale de Lausanne, 1015 Lausanne, Switzerland}
\affiliation{Dipartimento di Fisica, Università degli Studi di Milano, via Celoria 16, 20133 Milano, Italy}
\author{Davide Emilio Galli}
\affiliation{Dipartimento di Fisica, Università degli Studi di Milano, via Celoria 16, 20133 Milano, Italy}
\author{Mario Motta}
\affiliation{IBM Quantum, IBM Research Almaden, 650 Harry Road, San Jose, CA 95120, USA}
\author{Barbara Jones}
\affiliation{IBM Quantum, IBM Research Almaden, 650 Harry Road, San Jose, CA 95120, USA}

\begin{abstract}
The application of quantum algorithms to the study of many-particle quantum systems requires the ability to
 prepare wavefunctions that are relevant in the behavior of the system under study.
Hamiltonian symmetries are an important instrument, 
to classify relevant many-particle wavefunctions, and to improve the efficiency of numerical simulations.

In this work, quantum circuits for the exact and approximate preparation of total spin eigenfunctions on quantum computers are presented.
Two different strategies are discussed and compared: 
exact recursive construction of total spin eigenfunctions based on the addition theorem of angular momentum, 
and heuristic approximation of total spin eigenfunctions based on the variational optimization of a suitable cost function.
The construction of these quantum circuits is illustrated in detail, 
and the preparation of total spin eigenfunctions is demonstrated on IBM quantum devices, 
focusing on 3- and 5-spin systems on graphs with triangle connectivity.
\end{abstract}

\maketitle

\section{Introduction}

One of the central goals of quantum mechanics is to determine the behavior of multiple interacting particles. A fundamental and important example is the determination of the eigenstates of spin Hamiltonians, especially those with the possibilities of frustration. Of particular richness is the behavior of spin liquids \cite{wen2017colloquium,sachdev2018topological}, and signatures of spin-liquid behavior have been observed in a variety of magnetic materials \cite{fu2015evidence,banerjee2018excitations,anderson1987resonating}.

Quantum spin liquids are exotic phases of matter featuring topological order and long-range
quantum entanglement, which motivated the proposal and exploration of techniques to engineer such systems for topological protection of quantum information \cite{kitaev2003fault,nayak2008non}.
To gain a more profound understanding of quantum spin liquids requires concerted experimental and computational efforts.

Digital quantum computers have been proposed as an alternative and complementary approach to the 
exploration of these strongly correlated quantum phases \cite{georgescu2014quantum,bauer2020quantum}.
Systems with frustration caused by the lattice geometry or long-range interactions have been identified as promising avenues in the search for quantum spin liquids \cite{savary2016quantum,rokhsar1988superconductivity,sachdev1992kagome,misguich2002quantum,moessner2001resonating,read1991large,samajdar2021quantum}.
Spin liquid behavior in frustrated quantum systems has been explored with a variety of computational platforms, namely annealers \cite{zhou2020experimental}, superconducting quantum circuits \cite{song2018demonstration,andersen2020repeated}, and atom arrays \cite{semeghini2021probing}.

Hamiltonian symmetries play a central role in the classification of Hamiltonian eigenfunctions,
and in improving the efficiency and the accuracy of algorithms for the solution of the Schr\"{o}dinger equation. 
The incorporation of symmetries in quantum simulation algorithms is an active research area
with many important declinations, from qubit reduction \cite{bravyi2017tapering,setia2020reducing,faist2020continuous,elfving2021simulating,eddins2021doubling}, 
to careful design of state preparation circuits \cite{gard2020efficient}, 
to the introduction of symmetry violation penalties \cite{kuroiwa2021penalty}, 
to the adoption and refinement of post-selection and error-mitigation techniques.
An important example, particularly for spin Hamiltonians, is spin symmetry: when a Hamiltonian operator $\hat{H}$ commutes with the total spin and spin-$z$ operators,
respectively $\hat{S}^2$ and $\hat{S}_z$, there exists a basis of simultaneous eigenfunctions of $\hat{H}$ and such constants of motion.
Spin symmetry is encountered across a wide variety of situations relevant for the characterization of spin liquid behavior: 
from the Heisenberg model on a complete graph $\hat{H} = \sum_{i<j} \spin{i} \cdot \spin{j} = ( \hat{S}^2 - \sum_i \hat{S}_i^2 ) /2$,
to spin Hamiltonian providing a minimal model of magnetic correlations in molecular systems \cite{logemann2017exchange,schurkus2020exploring},
and the electronic structure of molecules and materials \cite{sugisaki2018quantum,rost2020simulation,jones2010spin,cao2019quantum,motta2021emerging}.

In this work, we explore quantum circuits for the synthesis of spin eigenfunctions on quantum computers.
Previous approaches have typically concentrated on circuits for exact or controllably approximated encoding of general spin eigenstates \cite{bacon2006efficient},
that require quantum resources (ancillae and multi-controlled multi-qubit gates) often beyond the capabilities of contemporary quantum hardware,
or circuits for encoding of specific spin eigenstates \cite{sugisaki2019open,bartschi2019deterministic}, 
that have less generality but require less quantum resources.

Here, we investigate the balance between generality, accuracy, and computational cost in the encoding of spin eigenfunctions by quantum circuits,
by pursuing two approaches: an exact recursive construction of spin eigenstates, and a heuristic variational construction of approximate spin eigenstates.

The remainder of the present work is structured as follows. 
In Section \ref{sec:methods} we present the two families of circuits studied in this work.
In Section \ref{sec:results} we apply the proposed circuits to systems of 3 and 5 spin-$1/2$ particles on graphs with triangle connectivities,
using classical simulators of quantum computers and quantum hardware.
Conclusions are drawn in Section \ref{sec:conclusions}, and additional theoretical and implementation details are presented in the Appendix.

\section{Methods}
\label{sec:methods}

In this Section, we describe our approaches to encode spin eigenstates on quantum computers.
The exact recursive construction is presented in Subsection \ref{sec:erc}, focusing on the structure and computational cost of the corresponding quantum circuits.
The heuristic variational construction is presented in Subsection \ref{sec:hvc}, focusing on the Ans\"{a}tze proposed and tested in the present work.

\subsection{Exact recursive construction}
\label{sec:erc}

In this Section, we consider a system of $n$ spin-$\frac{1}{2}$ particles, and our aim is to prepare such a system in an eigenfunction
$| \ell, m \rangle$ of the total spin operators $\hat{S}^2$ and $\hat{S}_z$ with eigenvalues {$\ell (\ell+1)$ and $m$ respectively, 
with $\ell=k/2$, $k \in \{1,2,\dots,n\}$ and $m = - \ell, \dots, \ell$.
Following an established convention, we map the Hilbert space of a single spin-up particle onto that of a single qubit as
\begin{equation}
| \ell=1/2 , m=1/2 \rangle \to | 0 \rangle \quad,\quad
| \ell=1/2 , m=-1/2 \rangle \to | 1 \rangle \quad.
\end{equation}
A single qubit can thus be prepared in a spin eigenfunction by applying to the initial state $|0 \rangle$ one of the following unitary transformations,
\begin{equation}
\label{eq:erc_0}
\hat{U}_{\ell=1/2, m=1/2} = \mathbbm{1} \quad,\quad
\hat{U}_{\ell=1/2, m=-1/2} = X \quad,
\end{equation}
where $X$ denotes the Pauli $X$ operator, defined in Appendix \ref{appendix:qis}.
A register of $n=2$ qubits can be prepared in a total spin eigenfunction using the quantum circuit
shown in Figure~\ref{fig:FIGURE_1}, which uses {a single ${\mathsf{CNOT}}$} gate,
also defined in Appendix \ref{appendix:qis}.

\begin{figure}[h!]
\includegraphics[width=0.15\textwidth]{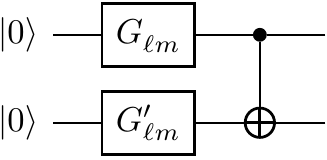}
\hspace{2cm}
\begin{tabular}{cccccc}
\hline\hline
$\ell$ & $m$ & $| \ell , m \rangle$ & $G_{\ell m}$ & $G_{\ell m}^\prime$ \\
\hline
0 &  0 & $\frac{|01\rangle-|10\rangle}{\sqrt{2}}$ & $HX$ & $X$ \\
1 &  1 & $|00\rangle$ & $\mathbbm{1}$ & $\mathbbm{1}$ \\
1 &  0 & $\frac{|01\rangle+|10\rangle}{\sqrt{2}}$ & $H$ & $X$ \\
1 & -1 & $|11\rangle$ & $X$ & $\mathbbm{1}$ \\
\hline\hline
\end{tabular}
\caption{Left: exact construction of total spin eigenfunctions for $n=2$ qubits.
Right: qubit representation of the total spin eigenfunction $|\ell , m \rangle$
for $n=2$ qubits, and gates $G_{\ell,m}$ used to encode it.}
\label{fig:FIGURE_1}
\end{figure}

In the remainder of this section, we will present a recursive procedure to prepare $n \geq 2$ qubits 
in a total spin eigenfunction. For $n=2$, such a preparation was {shown} in Figure~\ref{fig:FIGURE_1}. 

Let us now show that the possibility to prepare $n \geq 2$ qubits into a total spin eigenfunction implies the possibility to prepare $n+1$ qubits into a total spin eigenfunction.
The starting point is the angular momentum addition theorem,
\begin{equation}
\label{eq:tam1}
\ket{\ell_1 ,\ell_2, \ell  , m} = \sum_{m_1,m_2} C_{\ell_1, m_1, \ell_2, m_2}^{\ell, m} \ket{\ell_1 ,m_1} \otimes \ket{\ell_2 , m_2}
\end{equation}
where $\ket{\ell_1 , m_1}$, $\ket{\ell_2 , m_2}$, and $\ket{\ell_1 , \ell_2 , \ell  ,m}$ are eigenfunctions of the total spin operators with eigenvalues 
$\ell_1,m_1$, $\ell_2,m_2$ and $\ell,m$ respectively, and $C_{\ell_1, m_1, \ell_2, m_2}^{\ell, m}$ is a Clebsch-Gordan coefficient. 
Here, our goal is to map $\ket{\ell_1 , \ell_2, \ell  , m}$ onto a wavefunction of $n+1$ qubits. 
To this end, let us consider the case $\ell_2 = 1/2$, so that Eq.~\eqref{eq:tam1} takes the form
\begin{equation}
\label{eq:tam2}
\ket{\ell_1 , 1/2, \ell  , m} = \sum_{m_2} \psi_{m_2} \ket{\ell_1 ,m-m_2} \otimes \ket{ 1/2 , m_2} \quad,\quad
\psi_{m_2} = C_{\ell_1, m-m_2 , 1/2, m_2}^{\ell, m} \quad. \\
\end{equation}
By the induction hypothesis, $\ket{\ell_1 ,m-m_2}$ can be mapped onto an $n$-qubit wavefunction applying a quantum circuit $\hat{U}_{\ell_1,m-m_2}$
to a register of $n$ qubits prepared in $|0 \rangle^{\otimes n}$. Writing
\begin{equation}
\psi_{1/2} = \cos(\theta/2) 
\quad,\quad
\psi_{-1/2} = \sin(\theta/2) 
\quad
\end{equation}
for a suitable angle $\theta \in [0,2\pi)$ 
shows that the state $\ket{\ell_1 , 1/2, \ell  , m}$ can be mapped onto an 
$(n+1)$-qubit wavefunction applying {either of the the unitary transformations}
\begin{align}
\hat{U}_{\ell,m} 
&= \mathsf{c}_{n+1}\left( \hat{U}^{\phantom{\dagger}}_{\ell_1,m+1/2}  \hat{U}^\dagger_{\ell_1,m-1/2}  \right)
\hat{U}^{\phantom{\dagger}}_{\ell_1,m-1/2} \otimes R_{y}(\theta) \label{eq:tam3a} \\
&= \overline{\mathsf{c}}_{n+1} \hat{U}_{\ell_1,m-1/2} \, \mathsf{c}_{n+1} \hat{U}_{\ell_1,m+1/2} \, R_{y,n+1}(\theta) \label{eq:tam3b}
\end{align}
to a register of $n+1$ qubits prepared in $|0 \rangle^{\otimes (n+1)}$. In Eq.~\eqref{eq:tam3a} and~\eqref{eq:tam3b},
$R_{y,i}(\theta) = \mbox{exp}(-i \theta Y_i/2)$ is a $Y$ rotation of an angle $\theta$ applied to qubit $i$, and the symbols $\mathsf{c}_{i}$ and $\overline{\mathsf{c}}_{i}$ indicate
application of a unitary transformation when qubit $i$ is in $|0 \rangle$ or $|1 \rangle$ respectively.
{The unitary transformations in Eq.~\eqref{eq:tam3a} and~\eqref{eq:tam3b} are equivalent to the quantum circuits shown in Figure~\ref{fig:FIGURE_2}.}

\begin{figure}[h!]
\includegraphics[width=0.7\textwidth]{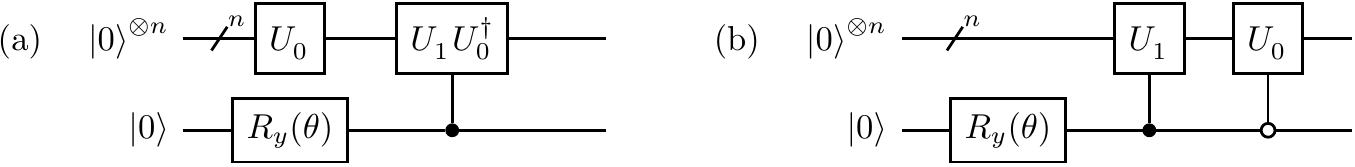}
\caption{Quantum circuits for exact recursive construction of total spin eigenfunctions, with
$\hat{U}_0 = \hat{U}_{\ell_1,m-1/2}$ and $\hat{U}_1 = \hat{U}_{\ell_1,m+1/2}$ to avoid clutter. The two circuits (a) and (b) are equivalent to each other, and Eq.~\eqref{eq:tam3a} and \eqref{eq:tam3b} respectively.}
\label{fig:FIGURE_2}
\end{figure}

As a simple and relevant example, in Figure~\ref{fig:FIGURE_3} we apply the exact recursive construction to $n=3$ spins. The gates $G_0$, $G_0^\prime$, $G_1$, $G_1^\prime$ are defined 
as in the right portion of Figure~\ref{fig:FIGURE_1},
and implement the unitaries $\hat{U}_{\ell_{01},m-1/2}$ and $\hat{U}_{\ell_{01},m+1/2}$ respectively,
with 
\begin{equation}
| \ell_{01} - 1/2 | \leq \ell \leq \ell_{01} + 1/2 \quad .
\end{equation}
Comparing Figures~\ref{fig:FIGURE_2}b and ~\ref{fig:FIGURE_3} shows that the latter circuit is simplified, in order to remove redundant gates.

\begin{figure}[h!]
\includegraphics[width=0.45\columnwidth]{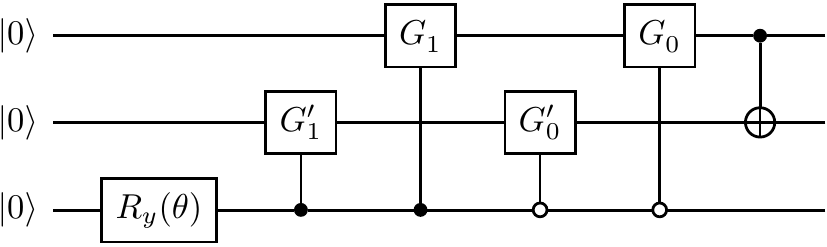}
\caption{Illustration of the exact recursive construction for $n=3$ spins. The single-qubit gates $G_k$ and $G^\prime_k$ are as listed in the right portion of Figure~\ref{fig:FIGURE_1}.}
\label{fig:FIGURE_3}
\end{figure}

\subsubsection{Computational cost}

Compared against previous approaches, the exact recursive construction presented here requires no ancillae \cite{bacon2006efficient}, and applies to generic spin eigenfunctions \cite{sugisaki2019open}.
However, it requires a number of single-qubit and ${\mathsf{CNOT}}$ gates scaling exponentially with $n$.

This circumstance can be understood considering the structure in Figure~\ref{fig:FIGURE_2}. Constructing an $(n+1)$-qubit total spin eigenfunction
requires two controlled $n$-qubit circuits and a single-qubit {$Y$ rotation}. 
Denoting $c_n$, $s_n$ the number of ${\mathsf{CNOT}}$ and single-qubit gates comprised
by the $n$-qubit circuit respectively, one can easily see that the $n(+1)$-qubit circuit requires $c_n$ Toffoli, $s_n$ controlled single-qubit gates, and 1
additional single-qubit gate.
By Lemmas 6.1 and 5.4 of \cite{barenco1995elementary} respectively, a Toffoli gate requires $8$ ${\mathsf{CNOT}}$ and 6 single-qubit gates, 
and a controlled single-qubit gate requires 2 ${\mathsf{CNOT}}$ and 2 single-qubit gates. 
Therefore, $c_{n+1} = 16 c_n + 4 s_n$ and $s_{n+1} = 12 c_n + 4 s_n$. 

Numeric integration of these recursion relations,
with the initial conditions $c_2, s_2 = 1,2$ as seen in Figure~\ref{fig:FIGURE_2}, gives $c_n, s_n \simeq 3^n$, which is exponential in the number of qubits.
We remark that, in practice, circuit simplification techniques can lower this gate count, 
especially for total spin eigenfunctions with low entanglement.

\subsection{Heuristic variational construction}
\label{sec:hvc}

The high computational cost of exact construction techniques for arbitrary spin 
eigenfunctions makes it desirable to propose {less expensive alternative strategies},
that are more suited to contemporary quantum devices. 
In this Section, we present a technique for heuristic variational construction 
of approximate total spin eigenfunctions.

Simulations of many-body ground- and excited-states on contemporary quantum hardware 
frequently make use of parametrized families of wavefunctions, 
of the form $| \Psi(\parameter{\theta}) \rangle = \hat{U}(\parameter{\theta}) | \Psi_0 \rangle$, where $\hat{U}(\parameter{\theta})$ is a quantum circuit comprising free parameters, 
and $|\Psi_0 \rangle$ is an initial wavefunction \cite{cerezo2020variational}. 
The best approximation to a target state within the set $| \Psi(\parameter{\theta}) \rangle$ is chosen by minimizing a suitable cost function with respect
to the parameters $\parameter{\theta}$ in the quantum circuit.
A prominent example is the variational quantum eigensolver \cite{peruzzo2014variational}, where the target state is the ground state of a Hamiltonian operator $\hat{H}$,
and the cost function is the expectation value $E(\parameter{\theta}) = \langle \Psi(\parameter{\theta}) | \hat{H} | \Psi(\parameter{\theta}) \rangle$ of the Hamiltonian over the 
state $| \Psi(\parameter{\theta}) \rangle$.
A natural choice for the cost function is 
\begin{equation}
\label{eqn:cost_function_a}
C_{\ell,m}(\parameter{\theta}) 
= \Big( \langle \hat{S}_z \rangle - m \Big)^2 
+ \Big( \langle \hat{S}^2 \rangle - \ell (\ell+1) \Big)^2 
\end{equation}
where $\langle \cdot \rangle$ denotes expectation value of an operator over the state $\Psi(\parameter{\theta})$.
In the present work, we also consider the possibility to fix the total spin of one or
more qubit subsets $\Omega_1, \dots, \Omega_r \subset \{ 0 \dots n-1 \}$ to a target value,
respectively $\ell_1, \dots, \ell_r$. To achieve this goal, we alter the cost function
as follows,
\begin{equation}
\label{eqn:cost_function}
C_{\ell_1, \dots, \ell_r , \ell, m}(\parameter{\theta}) 
= 
C_{\ell,m}(\parameter{\theta}) 
+ \sum_{k=1}^r \Big( \langle \hat{S}_{\Omega_k}^2 \rangle - \ell_k (\ell_k+1) \Big)^2 \quad,
\end{equation}
where and $\hat{S}_{\Omega_k}^2$ denotes the total spin of subset $\Omega_k$.
The functional $C_{\ell_1, \dots, \ell_r , \ell, m}(\parameter{\theta})$ takes value $0$ provided that $\Psi(\parameter{\theta})$ is an eigenfunction of the operators
$\hat{S}_{\Omega_k}^2$, $\hat{S}^2$, and $\hat{S}_z$ with eigenvalues $\ell_k (\ell_k+1)$,
$\ell (\ell+1)$, and $m$ respectively.

The gradient of the cost function Eq.~\eqref{eqn:cost_function} is computed using the chain
rule, and gradients of expectation values $\langle \cdot \rangle$ are computed analytically,
rather than by finite differences, using a suitable shift formula \cite{mcclean2016theory,parrish2019hybrid,schuld2019evaluating,mitarai2020theory,kottmann2021feasible}.

\subsubsection{Ans\"{a}tze}

The quality of a variational simulation depends not only on the cost function, but also on the underlying Ansatz,
i.e. the map $\parameter{\theta} \to | \Psi(\parameter{\theta}) \rangle$. In the present work, we explore several Ans\"{a}tze, with the goal
of assessing and comparing their accuracy and efficiently in the approximation of total spin eigenfunctions. 

First, we consider the following $R_y$ Ansatz \cite{kandala2017hardware},
\begin{equation}
\label{eq:ry}
| \Psi(\parameter{\theta}) \rangle 
= 
\prod_{k=1}^{r}
\left[
\prod_{i=0}^{n-1} R_{y,i}(\theta_{k,i}) E
\right]
\prod_{i=0}^{n-1} R_{y,i}(\theta_{0,i}) | \Psi_0 \rangle
\quad,\quad 
E = \prod_{(i,j) \in L} \mathsf{c}_iX_{j} \quad,
\end{equation}
where $| 0 \rangle^{\otimes n}$ is an initial wavefunction, $m$ is the number of qubits, 
$R_{y,i}(\theta) = \mbox{exp}(-i \theta Y_i/2)$ is a $Y$ rotation of an angle $\theta$ applied to qubit $i$, 
$\mathsf{c}_iX_{j}$ is a ${\mathsf{CNOT}}$ gate acting on qubits $(i,j) \in L$, and $r$ is an integer denoting the 
number of times an entangling gate $E$ followed by a layer of $Y$ rotations is repeated.
The $R_y$ Ansatz is illustrated in Figure~\ref{fig:FIGURE_4} for $r = 3$.

\begin{figure}[h!]
\includegraphics[width=0.75\columnwidth]{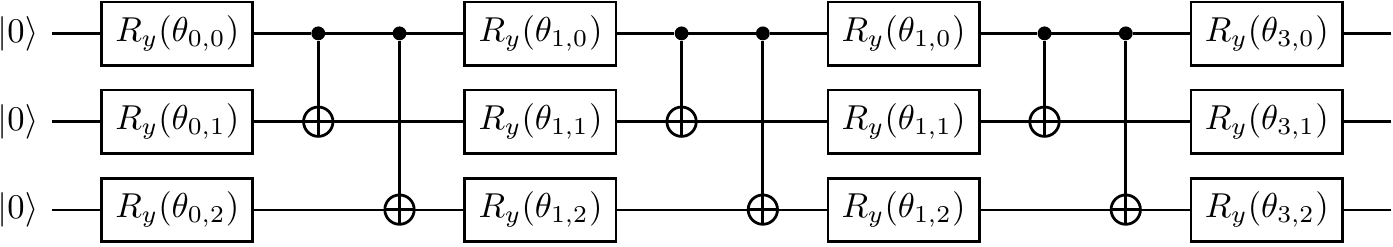}
\caption{Illustration of the $R_y$ Ansatz, with ${\mathsf{CNOT}}$ gates acting on qubits (0,1) and (1,2),
and initial state $|\Psi_0 \rangle = | 0,0,0 \rangle$. }
\label{fig:FIGURE_4}
\end{figure}

Next, we consider an Ansatz inspired by time evolution. 
Ref.~\cite{sharma2021preparation} illustrated a recursive procedure for the preparation of an $n$-qubit system in a total spin eigenfunction.
An $(n-1)$-qubit system is prepared in a total spin eigenfunction $| \ell , m \rangle_{n-1}$ and coupled with a qubit prepared in $| 1/2,1/2 \rangle$.
The system evolves under the action of a Heisenberg Hamiltonian with all-to-all coupling, $\hat{H} = J \sum_{i<j=0}^{n-1} {\bf{S}}_i \cdot {\bf{S}}_j$,
leading to the wave-function
\begin{equation}
| \Psi(t) \rangle 
= e^{ - i t \hat{H} } | \ell , m \rangle_{n-1} \otimes | 1/2,1/2 \rangle 
= \sum_{k=\pm 1} c_k(t) | \ell+k/2 , m+1/2 \rangle
\quad.
\end{equation}
Provided that the coefficients satisfy the relation
\begin{equation}
c_+(t) = \sqrt{ \frac{\ell + m + 1}{2 \ell +1} } 
\quad,\quad
c_-(t) = \sqrt{ \frac{\ell - m}{2 \ell +1} }
\quad,
\end{equation}
the state $| \Psi(t) \rangle$ is a total spin eigenfunction with eigenvalues $\ell$, $m$.
Motivated by this results, following and adapting the construction of the QAOA method \cite{farhi2014quantum}, 
we introduce a {variational form inspired by time evolution}, shown in Figure~\ref{fig:FIGURE_5}.
The Hamiltonian is represented as 
\begin{equation}
\hat{H} = J \sum_{i<j=0}^{n-1} {\bf{S}}_i \cdot {\bf{S}}_j = J \sum_{i<j=0}^{n-2} {\bf{S}}_i \cdot {\bf{S}}_j + J \left[ \sum_{i=0}^{n-2} {\bf{S}}_i \right] \cdot {\bf{S}}_{n-1} = \hat{H}_0 + \hat{V} \quad,
\end{equation}
and in Figure~\ref{fig:FIGURE_5}a the time evolution operator is approximated with the following primitive Trotter formula \cite{trotter1959product,suzuki1976generalized,childs2019theory,childs2019nearly},
\begin{equation}
e^{-it \hat{H} } \simeq e^{-it \hat{V} } e^{-it \hat{H}_0 } \quad.
\end{equation}
In Figure~\ref{fig:FIGURE_5}b, we further impose the following primitive Trotter approximation, 
\begin{equation}
e^{ - i tJ \sum_{ij}  {\bf{S}}_i \cdot {\bf{S}}_j } \simeq \prod_{ij} e^{ - i tJ {\bf{S}}_i \cdot {\bf{S}}_j } = \prod_{ij} u_{ij}(tJ)
\quad,
\end{equation}
and substitute each unitary $e^{ - i tJ {\bf{S}}_i \cdot {\bf{S}}_j }$ with a parametrized two-qubit gate of the kind shown in Figure~\ref{fig:FIGURE_5}c.

\begin{figure}[h!]
\includegraphics[width=0.9\columnwidth]{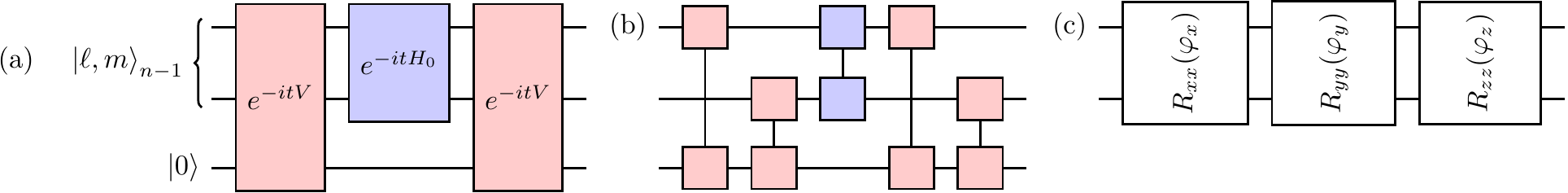}
\caption{Time evolution-inspired variational form. (a) time evolution under the Heisenberg Hamiltonian for $r=2$ steps and $n=3$ qubits is approximated with a primitive Trotter formula;
in the first step of time evolution, the unitary $e^{-it \hat{H}_0 }$ is not applied in the first step of time evolution, since the top $n-1$ qubits are prepared in an eigenstate of $\hat{H}_0$.
(b) each unitary $e^{-it \hat{H}_0 }$ and $e^{-it \hat{V} }$ (blue, red blocks in the left panel) is approximated by a product of unitary $u_{ij}(tJ)$ (rectangles connected by a vertical line).
(c) each unitary $u_{ij}(tJ)$ is replaced by a parametrized two-qubit gate.
}
\label{fig:FIGURE_5}
\end{figure}

\section{Results}
\label{sec:results}

\paragraph{Computational details.}
We use IBM's open-source Python library for quantum computing, Qiskit \cite{aleksandrowicz2019qiskit}. 
Qiskit provides tools for various tasks such as creating quantum circuits, performing simulations, and computations on quantum devices.
It also contains an implementation of the VQE algorithm, 
a hybrid quantum-classical algorithm that uses both quantum and classical resources 
to variationally optimize the cost function, and a classical exact eigensolver algorithm, against which to compare results.
We then minimize the expectation value of the cost function with respect to the parameters of the variational circuit. The minimization is carried out through a classical optimization method \cite{kingma2014adam,stokes2020quantum,spall1998overview,spall2000adaptive,s2013stochastic,hestenes1952methods}.
In this work, we used the conjugate gradient \cite{hestenes1952methods} on the noiseless and deterministic \simulator{statevector} of Qiskit. 
We then use the optimized parameters to compute expectation values on quantum hardware.
We performed experiments on 5-qubit devices available through IBM Quantum Experience, 
each with a Quantum Volume \cite{cross2019QV} of 32, namely, \device{athens}, 
\device{santiago}, and \device{manila} \cite{IBMQDevices}. 

\paragraph{Error mitigation.}
We employed readout-error mitigation (EM) \cite{temme2017error,kandala2019error,bravyi2020mitigating,maciejewski2020mitigation} as implemented in Qiskit Ignis to correct measurement errors. 
We also used a noise extrapolation scheme, {namely a simplified form of the} Richardson extrapolation (RE), that uses additional ${\mathsf{CNOT}}$ gates at the minimum-energy VQE iterations to account for errors introduced by two-qubit entangling operations \cite{dumitrescu2018cloud,stamatopoulos2019option}.
More specifically, each ${\mathsf{CNOT}}$ gate in the circuit is replaced with a product of $2k+1$ ${\mathsf{CNOT}}$ gates, $\mathsf{c}X \to \mathsf{c}X^{2k+1}$,
and the expectation value $A_k$ of an observable is extrapolated with a linear Ansatz, 
\begin{equation}
A_k = A_{\mathrm{extrapolated}} + (2k+1) \Delta \quad .
\end{equation}
Following Ref.~\cite{dumitrescu2018cloud}, we call
$r=2k+1$ {the "noise parameter"}.

\paragraph{Tomography.} An important goal of the present work 
is to assess how faithfully a quantum circuit executed on a quantum device
approximates a target wavefunction $\psi_\text{ideal}$.
To answer this question, we compute the fidelity
\begin{equation}
\label{eqn:fid_exp}
F(\rho_\text{circuit},\psi_\text{ideal}) = \langle \psi_\text{ideal}| \rho_\text{circuit} | \psi_\text{ideal} \rangle
\end{equation}
between $\psi_\text{ideal}$ and the density operator $\rho_\text{circuit}$ 
prepared on the quantum hardware, and the purity of the former,
\begin{equation}
\label{eqn:pur_exp}
P(\rho_\text{circuit}) = \mbox{Tr}( \rho^2_\text{circuit} ) \quad .
\end{equation}
The density operator $\rho_\text{circuit}$ is measured by quantum tomography \cite{samach2021lindblad}, and Eq.~\eqref{eqn:fid_exp} and \eqref{eqn:pur_exp}
are computed on the \simulator{statevector} of Qiskit based on the measurement
outcomes.

\paragraph{Target problems.} In the present work, 
we consider systems of $n=3$ and $n=5$ spins on graphs with triangle connectivity, illustrated in Figure~\ref{fig:FIGURE_6}.
For 3-spin problems, our goal is to prepare total spin eigenfunctions 
starting from total spin eigenfunctions 
of the left spins 
(labeled 0,1 in Figure~\ref{fig:FIGURE_6}a).
For 5-spin problems, we start from total spin eigenfunctions 
of the left spins 
(labeled 0,1 in Figure~\ref{fig:FIGURE_6}b and henceforth L),
of the left and central spins 
(labeled 0,1,2 in Figure~\ref{fig:FIGURE_6}b and henceforth LC), and 
of the right spins 
(labeled 3,4 in Figure~\ref{fig:FIGURE_6}b and henceforth R).

\begin{figure}[h!]
\centering
\includegraphics[width=0.45\textwidth]{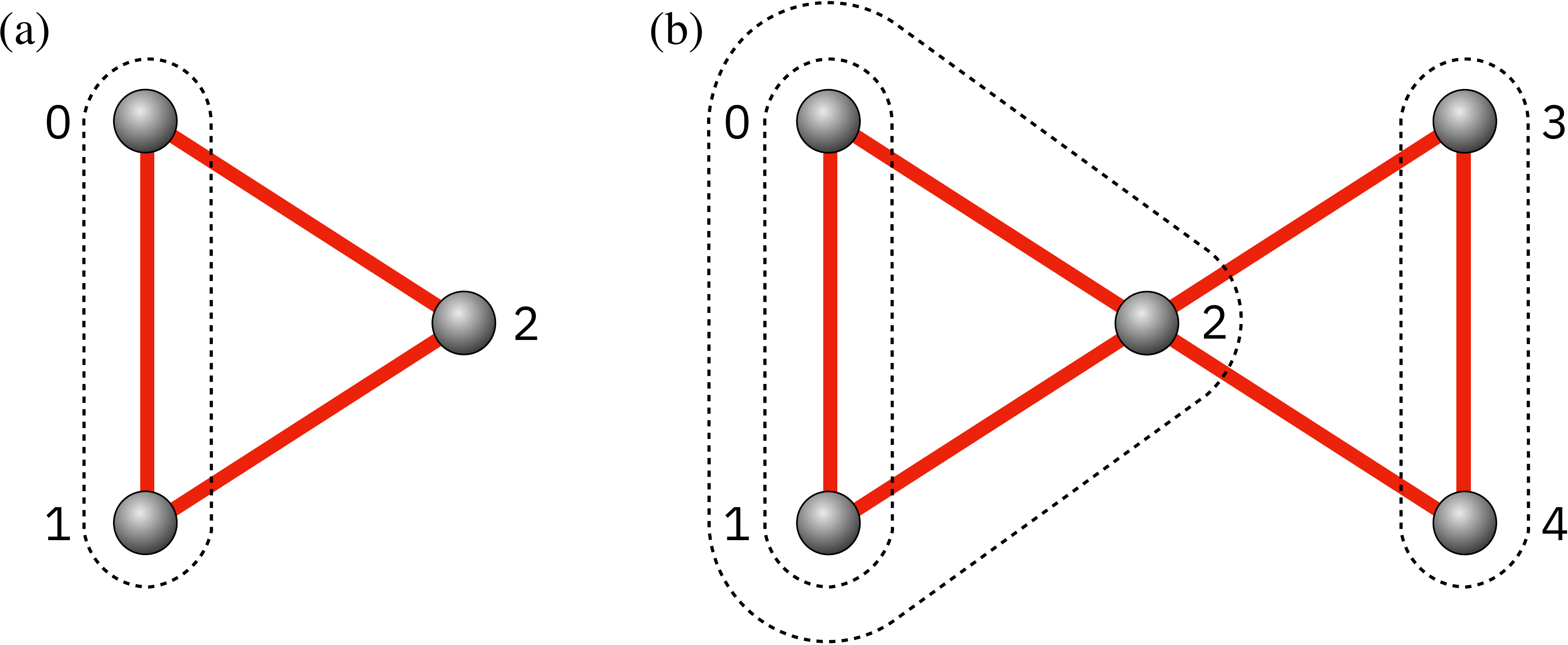}
\caption{Representation of a set of 3 (a) and 5 (b) spins
on graphs with triangle connectivity.
Black circles represent spins, labeled from 0 to 4, 
and red lines connect neighboring qubits.
Black dashes lines represent subsets $\Omega_{01}$ ({in panel} a) and $\Omega_L = \Omega_{01}$, $\Omega_{LC} = \Omega_{012}$, $\Omega_R = \Omega_{34}$ ({in panel} b)}
\label{fig:FIGURE_6}
\end{figure}

\subsection{Exact recursive construction}

In Figure \ref{fig:FIGURE_7} we demonstrate the exact recursive construction for systems of $n=3$ spins prepared in the representative 
states $\ket{\ell_{01}=0,\ell=1/2,m=-1/2}$ and $\ket{\ell_{01}=1,\ell=3/2,m=-1/2}$. We focus on the evaluation of the total spin operators
$\hat{S}_{01}^2$ (total spin of qubits 0 and 1), $\hat{S}^2$, and $\hat{S}_z$.
As naturally expected, the use of EM and EM+RE significantly improves 
the agreement between exact and simulated values for these quantities.

\begin{figure}[h!]
\centering
\includegraphics[width=0.8\textwidth]{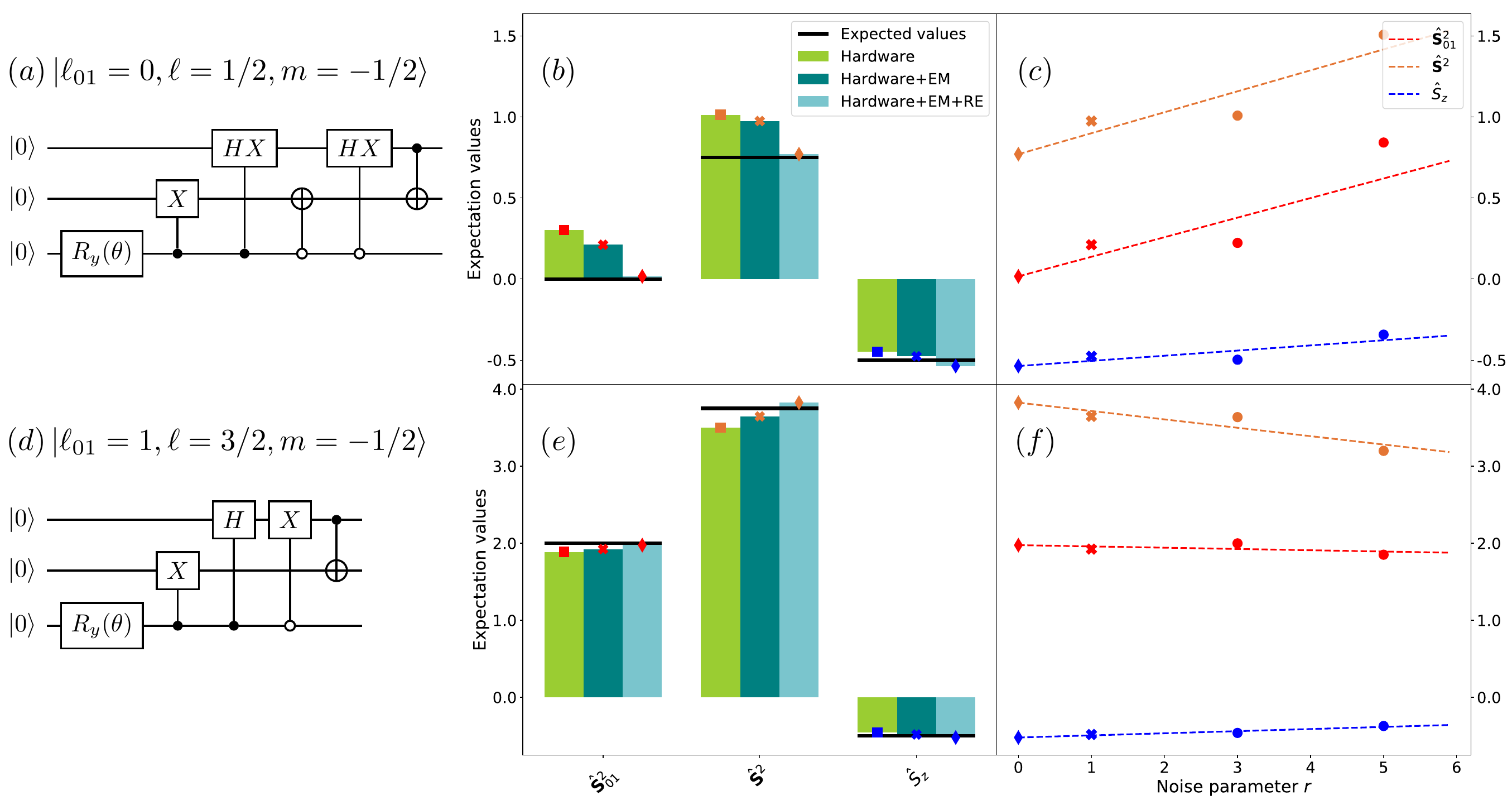} 
\caption{Results from the exact recursive construction, 
for systems of $n=3$ spins prepared in 
$\ket{\ell_{01}=0,\ell=1/2,m=-1/2}$ and $\ket{\ell_{01}=1,\ell=3/2,m=-1/2}$ 
((7a,b,c), (7d,e,f) respectively), on the \texttt{ibmq}$\_$\texttt{athens} device.
(7a,d): quantum circuits to encode the target states.
(7b,e): expectation values of the total spin operators $\hat{S}_{01}^2$, $\hat{S}^2$, and $\hat{S}_z$ from 
raw hardware simulations (light green, leftmost bar), 
simulations employing readout error mitigation (EM, teal, center bar) and 
simulations employing readout error mitigation and Richardson extrapolation (EM+RE, light blue, rightmost bar). Exact values are marked by black lines.
(7c,f): extrapolation of the expectation values in the middle panels versus 
number of CNOT gates. Crosses (r=1), circles (r=3,5), and diamonds (r=0) denote error-mitigated, noise-amplified, and extrapolated quantities respectively. 
Red, yellow, and blue symbols indicate $\hat{S}_{01}^2$, 
$\hat{S}^2$, and $\hat{S}_z$ respectively.
}
\label{fig:FIGURE_7}
\end{figure}

Two natural questions are whether the improved quality observed for expectation values of spin operators translates to generic observables,
and what decoherence mechanisms are responsible for the deviations seen in raw hardware simulations.

To address these questions, in Table \ref{tab:TABLE_1} we compute the 
fidelity and purity of the state $\rho_\text{circuit}$, respectively Eq.~\eqref{eqn:fid_exp} and \eqref{eqn:pur_exp}.
We observe $F(\rho_\text{circuit},\psi_\text{ideal})<1$, which mirrors the deviations between exact and simulated expectation values of spin operators.
Furthermore, $P(\rho_\text{circuit})<1$ in raw hardware simulations, indicating that qubits are depolarized by decoherence.
The combined use of EM+RE increases the fidelity towards 1, indicating a more faithful representation of the quantities measured on the device. Fidelities are extrapolated as shown in the right panel of Figure~\ref{fig:FIGURE_8}, where
analogous results are presented for all the total spin eigenfunctions of $n=3$ qubits.

\begin{table}[h!]
\centering
\begin{tabular}{ccccc}
\hline\hline
$(\ell_{01},\ell,m)$ & $F(\rho_\text{circuit},\psi_\text{ideal})$ (raw) & $P(\rho_\text{circuit})$ (raw) & $F(\rho_\text{circuit},\psi_\text{ideal})$ (EM+RE) \\
\hline
$(0,1/2,-1/2)$ & $0.78708$ & $0.79982$ & $0.9720 \pm 0.0306$ \\
$(1,3/2,-1/2)$ & $0.86417$ & $0.90659$ & $0.9944 \pm 0.0057$ \\
\hline\hline
\end{tabular}
\caption{Fidelity and purity of the state $\rho_\text{circuit}$, for systems of $n=3$ spins prepared in the states $\ket{\ell_{01}=0,\ell=1/2,m=-1/2}$ and $\ket{\ell_{01}=1,\ell=3/2,m=-1/2}$ (top, bottom respectively) using exact recursive construction. 
The fidelity is extrapolated as shown in the right panel of Figure~\ref{fig:FIGURE_8}.}
\label{tab:TABLE_1}
\end{table}

\begin{figure}[h!]
\centering
\includegraphics[width=0.7\textwidth]{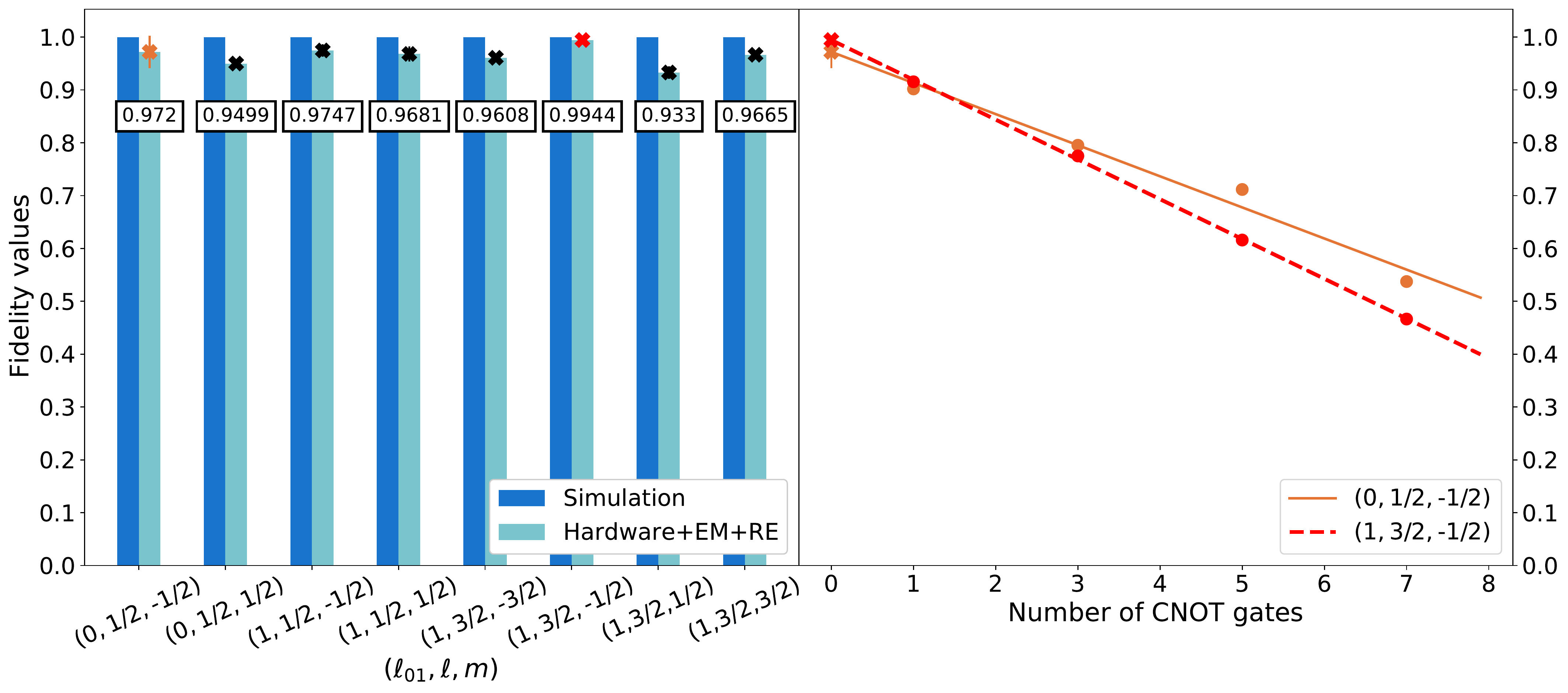}
\caption{
Left:
fidelities of $n=3$ qubits prepared in the total spin eigenstates  $\ket{\ell_{01},\ell,m}$. Fidelities are computed through qubit 
state tomography employing readout error mitigation (EM),
and gate noise is mitigated by a Richardson extrapolation (RE).
Right:
Richardson extrapolation of the error-mitigated fidelities for systems of $n=3$ qubits prepared in the states $\ket{\ell_{01}=0,\ell=1/2,m=-1/2}$ (yellow symbols) and $\ket{\ell_{01}=1,\ell=3/2,m=-1/2}$ (red symbols) with the exact recursive procedure.
Circles and crosses denote measured an extrapolated quantities respectively.
} 
\label{fig:FIGURE_8}
\end{figure}

In Figure~\ref{fig:FIGURE_9}, we apply the exact recursive construction to systems of $n=5$ spins, 
using the \simulator{qasm} with noise model from \device{athens}, 
and \device{athens}. We focus on the states $\ket{\ell_{L}=0,\ell_{LC}=1/2,\ell_R=0,\ell=1/2,m=-1/2}$ and $\ket{\ell_{L}=0,\ell_{LC}=1/2,\ell_R=1,\ell=1/2,m=-1/2}$.
Although the trends seen for $n=3$ qubits, 
especially the beneficial impact of EM+RE techniques, 
are confirmed, a significant difference exists 
between Figures \ref{fig:FIGURE_7} and \ref{fig:FIGURE_9}:
in the latter case, error mitigation techniques do not close the 
gap between exact and simulated expectation values.
The same phenomenon is observed in Figure~\ref{fig:FIGURE_10},
where we compute the fidelities between exact and simulated total spin eigenfunctions, for several representative states, 
using \simulator{qasm} and \device{manila}.
The different efficacy of error mitigation techniques for $n=3$
and $n=5$ qubits is an important observation of the present work,
and it will be discussed in detail in the conclusions.

\begin{figure}[h!]
\centering
\includegraphics[width=0.9\textwidth]{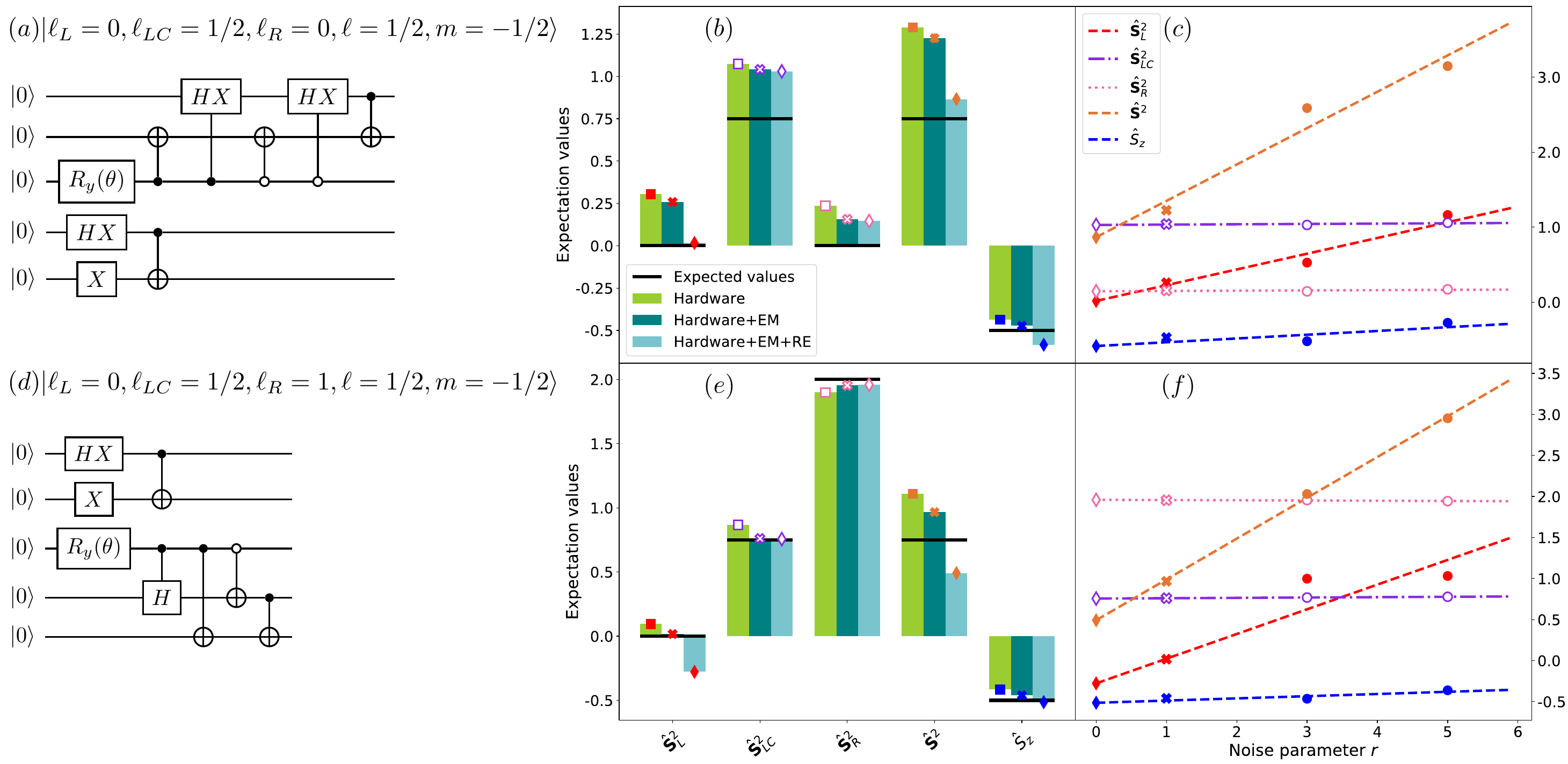}
\caption{
Results from the exact recursive construction, 
for systems of $n=5$ spins prepared in the wavefunctions
$\ket{s_{L}=0,s_{LC}=1/2,s_R=0,s=1/2,s_z=-1/2}$ and $\ket{s_{L}=0,s_{LC}=1/2,s_R=1,s=1/2,s_z=-1/2}$ 
((9a,b,c), (9d,e,f) respectively), on the \texttt{ibmq}$\_$\texttt{athens} device.
(9a,d): quantum circuits to encode the target states.
(9b,e): expectation values of the total spin operators $\hat{S}_{01}^2$, $\hat{S}^2$, and $\hat{S}_z$ from 
raw hardware simulations (light green, leftmost bar), 
simulations employing readout error mitigation (EM, teal, center bar) and 
simulations employing readout error mitigation and Richardson extrapolation (EM+RE, light blue, rightmost bar). Exact values are marked by black lines.
(9c,f): extrapolation of the expectation values in the middle panels versus 
number of CNOT gates. Crosses (r=1), circles (r=3,5), and diamonds (r=0) denote error-mitigated, noise-amplified, and extrapolated quantities respectively. 
Red, yellow, and blue symbols indicate $\hat{S}_{01}^2$, 
$\hat{S}^2$, and $\hat{S}_z$ respectively.
}
\label{fig:FIGURE_9}
\end{figure}

\begin{figure}[h!]
\centering
\includegraphics[width=0.45\textwidth]{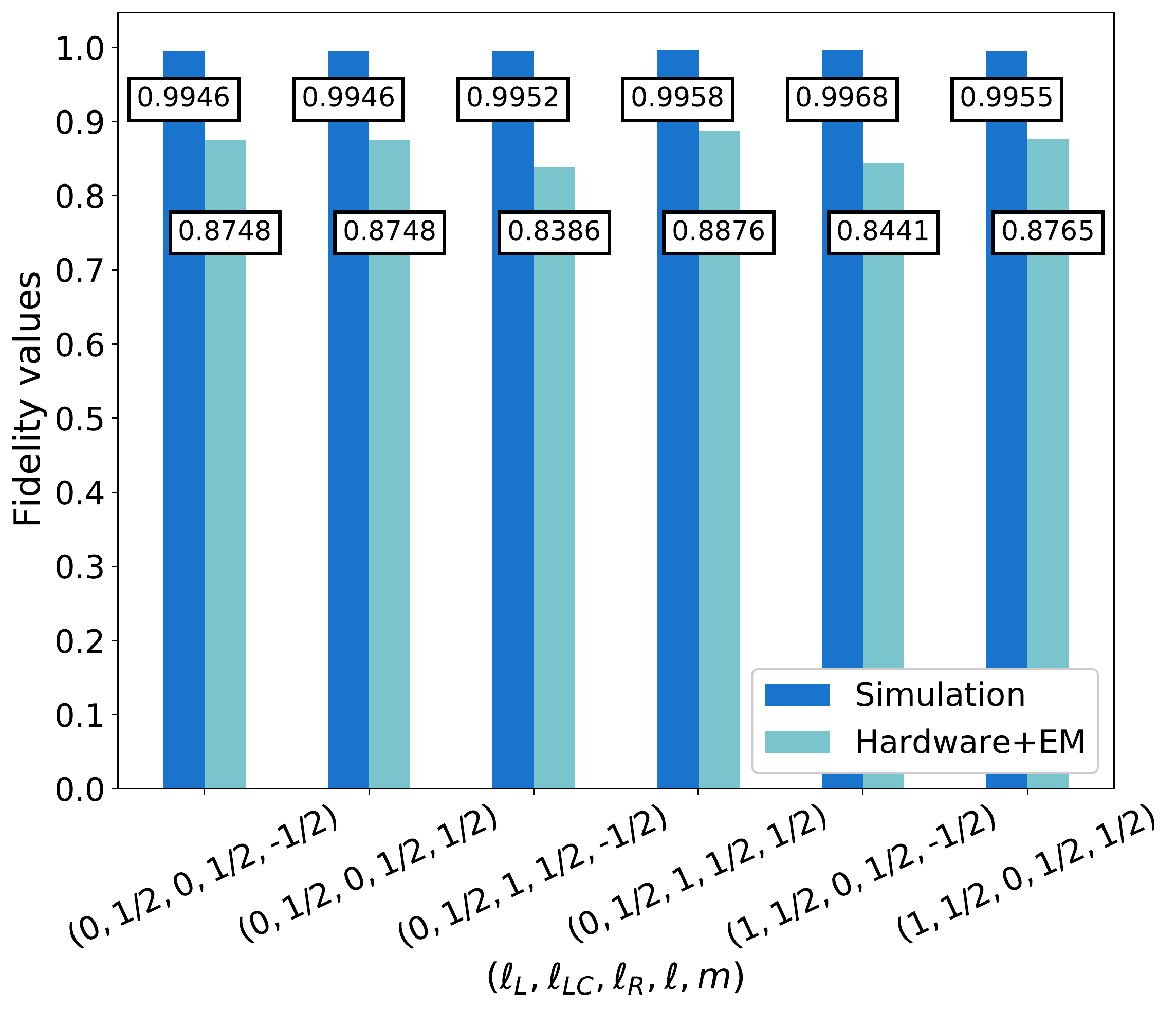}
\caption{Fidelities between exact and simulated total spin eigenfunctions for $n=5$ qubits, computed using \simulator{qasm} and \device{manila}.}
\label{fig:FIGURE_10}
\end{figure}

\subsection{Heuristic variational construction}

In this Section, we assess the accuracy and performance of VQE-based quantum circuits for approximate encoding of total spin eigenstates.
Unlike in the case of exact recursive construction, variationally optimized circuits are not guaranteed to accurately approximate all the total spin eigenstates.
For each Ansatz explored in the present work, we thus begin our analysis by assessing the Ansatz accuracy using classical simulators of noiseless quantum hardware,
and we finally demonstrate the corresponding quantum circuits on IBM devices.

\subsubsection{$R_y$ Ansatz}

\begin{figure}[h!]
\includegraphics[width=0.45\textwidth]{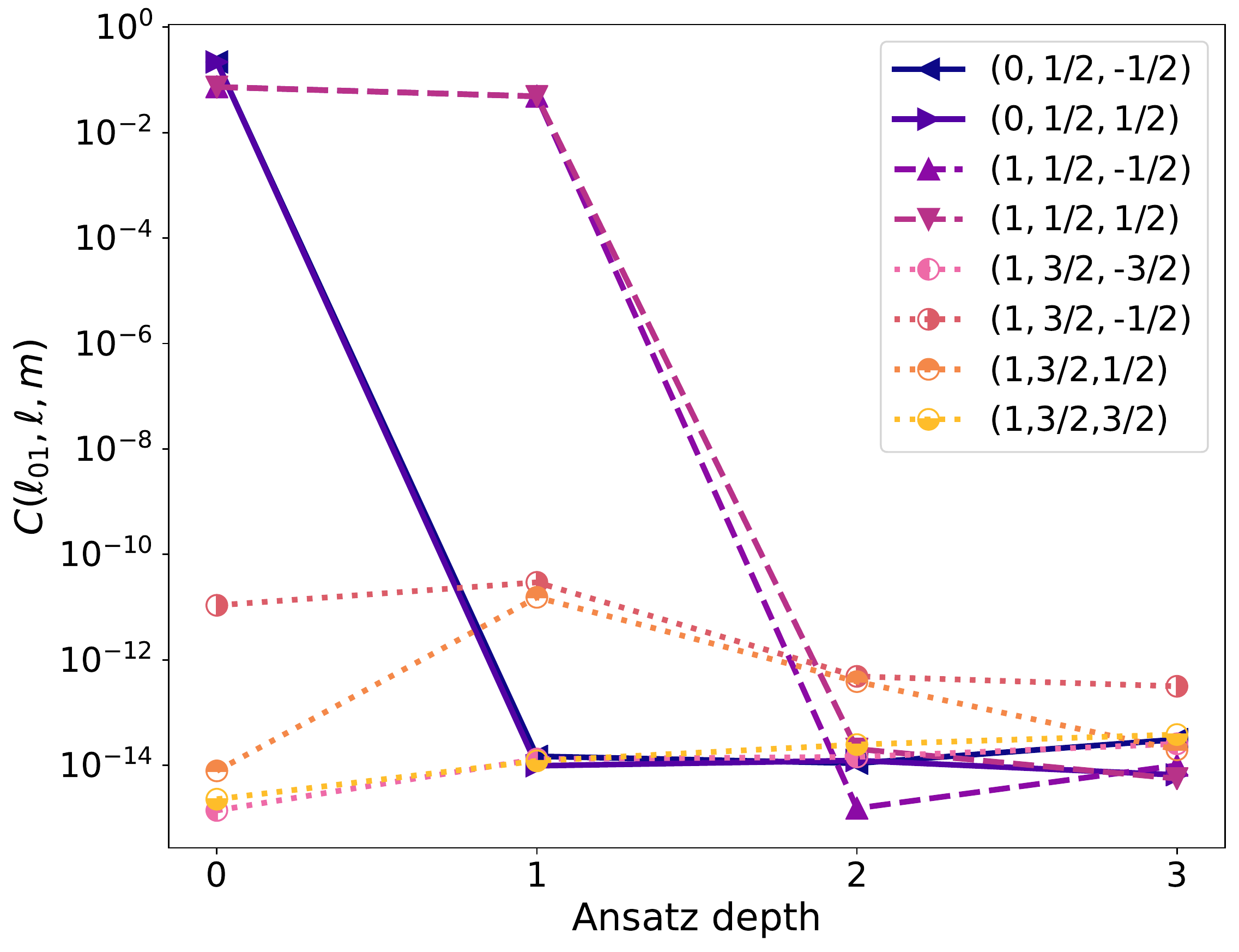}
\caption{Cost function values $C_\text{opt}$ of optimal quantum circuits for $n=3$ qubits using the $R_y$ Ansatz with depth $r=3$, computed with \simulator{statevector}. The ideal value of the cost function is zero.
}
\label{fig:FIGURE_11}
\end{figure}

\begin{table}[h]
\centering
\begin{tabular}{cc}
\toprule
$(\ell_{01},\ell,m)$ & $| \braket{ \psi_\text{ideal} | \Psi_\text{opt} } |^2$ \\
\hline
$(0,1/2,-1/2)$ & $1.000000$ \\
$(0,1/2,\phantom{-}1/2)$ & $1.000000$ \\
$(1,1/2,-1/2)$ & $1.000000$ \\
$(1,1/2,\phantom{-}1/2)$ & $1.000000$ \\
$(1,3/2,-3/2)$ & $1.000000$ \\
$(1,3/2,-1/2)$ & $0.396621$ \\
$(1,3/2,\phantom{-}1/2)$ & $0.024675$ \\
$(1,3/2,\phantom{-}3/2)$ & $1.000000$ \\
\hline
\end{tabular}
\caption{Fidelities of optimal quantum circuits for $n=3$ qubits using the $R_y$ Ansatz with depth $r=3$, computed with \simulator{statevector}.}
\label{tab:TABLE_2}
\end{table}

\begin{figure}[h!]
\includegraphics[width=0.9\textwidth]{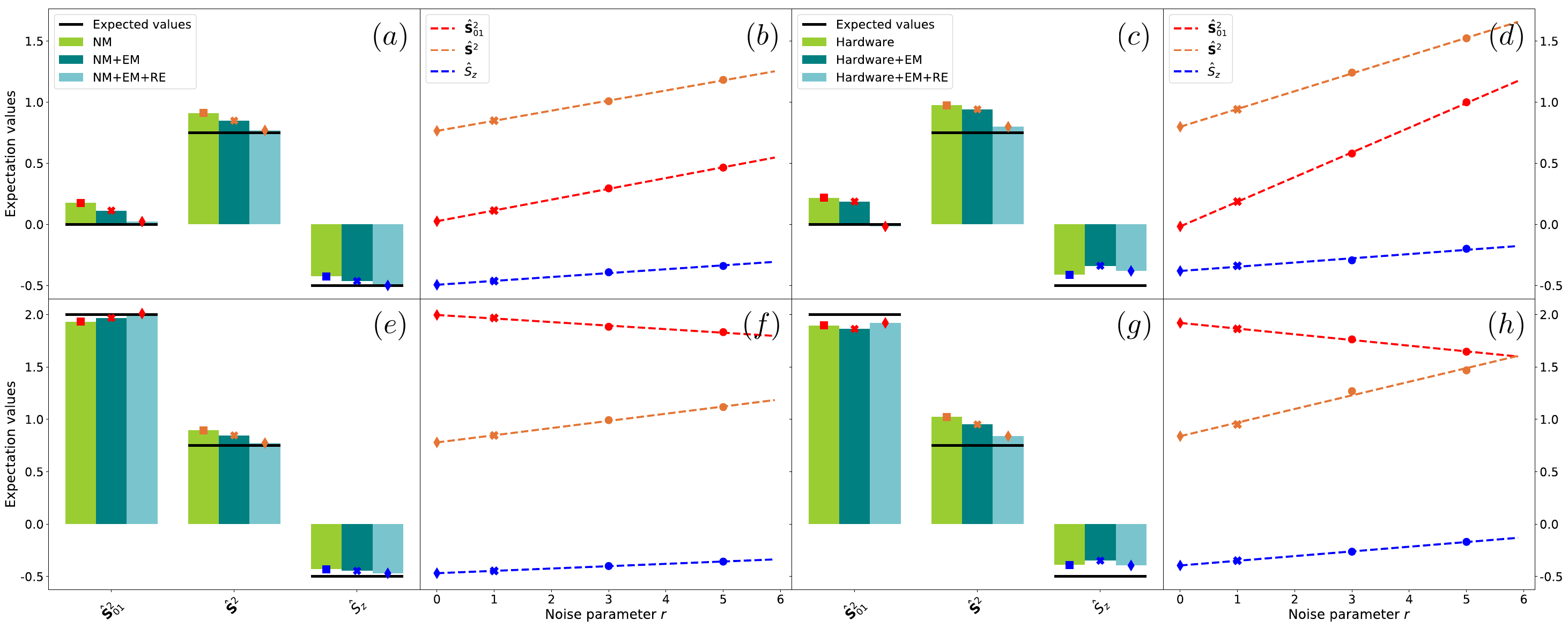}
\caption{Bar charts: expectation values of the spin operators $\hat{S}^2$, $\hat{S}_z$ and $\hat{S}_{01}^2$ for systems of $n=3$ qubits prepared in $\ket{0,1/2,-1/2}$ (12 a,c) and $\ket{1,1/2,-1/2}$ (12 e,g)
using \simulator{qasm} with noise model from \device{athens} (12 a,e) and on the device \device{athens} (12 c,g) and the $R_y$ Ansatz with depth $r=3$.
(12 b,d,f,h): Richardson extrapolation analysis for the expectation values shown in the bar charts.The vertical scales are matched to the scales in the accompanying bar chart.
}
\label{fig:FIGURE_12}
\end{figure}

\begin{figure}[h!]
\centering
\includegraphics[width=0.7\textwidth]{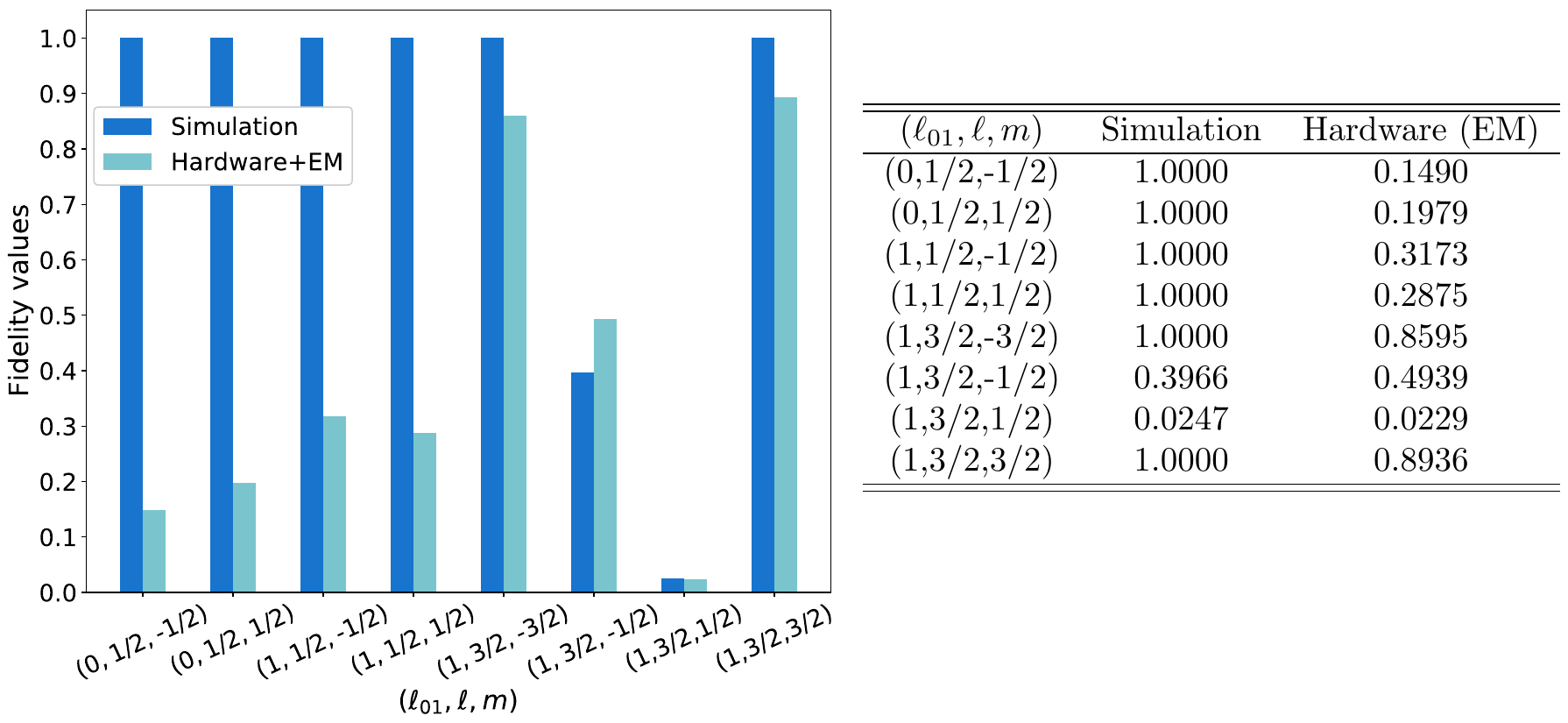}
\caption{Fidelities for $n=3$ between exact total spin eigenstates and VQE wavefunctions computed with \simulator{statevector} and \device{athens} (light, dark blue columns).}
\label{fig:FIGURE_13}
\end{figure}

In Figure~\ref{fig:FIGURE_11}, we show the optimized cost function $C_\text{opt}(\ell_{01},\ell,m) = \min_{\parameter{\theta}} C_{\ell,m,\ell_{01}}(\parameter{\theta})$ for the $R_y$ Ansatz with $n=3$ qubits {and the connectivity in Figure~\ref{fig:FIGURE_4}}, 
as a function of the Ansatz depth $r$, as defined in Eq.~\eqref{eq:ry}. As seen, the choice $r \geq 2$ ensures minimization of the cost function within single numeric precision across all spin eigenstates.
In Table~\ref{tab:TABLE_2}, we list the fidelity between the state $\Psi_\text{opt} = \Psi(\parameter{\theta}_\text{opt})$ and the target eigenstate $|\ell_{01},\ell,m\rangle$. As seen, the fidelity is equal
to 1 except for the Dicke states $\ket{1,3/2,\pm 1/2}$ \cite{dicke1954coherence,prevedel2009experimental,ozdemir2007necessary,toth2012multipartite,childs2000finding}.
This is not unexpected, as the Dicke states are the most entangled $n$-spin states, and therefore the most difficult targets for a variational optimization.

In Figure~\ref{fig:FIGURE_12} we compute the spin operators $\hat{S}^2$, $\hat{S}_z$ and $\hat{S}_{01}^2$ using \simulator{qasm} with noise model from \device{athens} and on the device \device{athens},
and the $R_y$ Ansatz with depth $r=3$. Similar trends are seen in both cases, with the combined use of EM+RE systematically improving results. 
It is clear that the {noise models do not faithfully emulate the noise observe in our experiments}, as readily seen when comparing the data in Figure~\ref{fig:FIGURE_12}.
Given that the results are highly sensitive to noise affecting the qubits, it will be important to investigate these noise source further and determine the extent to which they can mitigated.

While the combination of EM and RE reduces the deviations between exact and simulated quantities, it does not completely remove them. To document this observation, in Figure~\ref{fig:FIGURE_13}
we compute the fidelity between optimal exact total spin eigenstates and VQE wavefunctions computed with \simulator{statevector} (seen in Table~\ref{tab:TABLE_2}) and on \device{athens}.
As seen, decoherence significantly affects the fidelities of the obtained wavefunctions.

\subsubsection{Time-evolution variational form}

We now consider the time-evolution variational form. Unlike in the case of $R_y$, where the initial state is customarily fixed to $|0\rangle^{\otimes n}$, here the initial state is set to
\begin{equation}
\label{eqn:initial_states}
\begin{array}{ll}
|\psi(0) \rangle =  |110 \rangle                                                         & \mbox{for} \quad \ket{1,3/2,-1/2} \\
|\psi(0) \rangle =  \frac{|01\rangle + |10\rangle}{\sqrt{2}} |0 \rangle & \mbox{for} \quad \ket{1,3/2,1/2} \\
\end{array}
\end{equation} 

For $r \geq 2$ repetitions, the time-evolution variational form returned $C_\text{opt} < 10^{-18}$ and fidelity $F > 0.9999$ for all total spin eigenfunctions, including Dicke states.
It should be noted, on the other hand, that to the better accuracy of the time-evolution variational form corresponds a higher computational cost, quantified in Table~\ref{tab:TABLE_3}.

\begin{table}[h!]
\centering
\begin{tabular}{lcc}
\hline\hline
$(\ell_{01},\ell,m)$ & $d_{\simulator{statevector}}$ & $d_{\device{santiago}}$ \\
\hline
$(0,1/2,-1/2)$ & $76$ & $117$ \\
$(0,1/2,\phantom{-}1/2)$  & $76$ & $118$ \\
$(1,1/2,-1/2)$ & $76$ & $120$ \\
$(1,1/2,\phantom{-}1/2)$  & $76$ & $118$ \\
$(1,3/2,-3/2)$ & $76$ & $117$ \\
$(1,3/2,-1/2)$ & $76$ & $119$ \\
$(1,3/2,\phantom{-}1/2)$  & $76$ & $126$ \\
$(1,3/2,\phantom{-}3/2)$  & $65$ & $94$  \\
\hline\hline
\end{tabular}
\caption{Circuit depth required to approximate $3$-qubit total spin eigenstates using the time-evolution variational form with $r=2$ repetitions
from \simulator{statevector} (full connectivity, single-qubit and CNOT native gates) and \device{santiago} (linear connectivity, device-specific native gates).}
\label{tab:TABLE_3}
\end{table}

In Figure~\ref{fig:FIGURE_14} we show the expectation values of the total spin operators for $n=3$ qubits, 
using the time-evolution variational form, computed with \simulator{statevector} (seen in Table~\ref{tab:TABLE_2}) and on \device{santiago}.
The higher computational cost of the time-evolution variational form, compared with the $R_y$ variational form, results in an increased sensitivity to noise,
and ultimately in a worse agreement between exact and simulated quantities, a trend especially pronounced on the actual quantum device. 

\begin{figure*}[h!]
\centering
\includegraphics[width=0.9\textwidth]{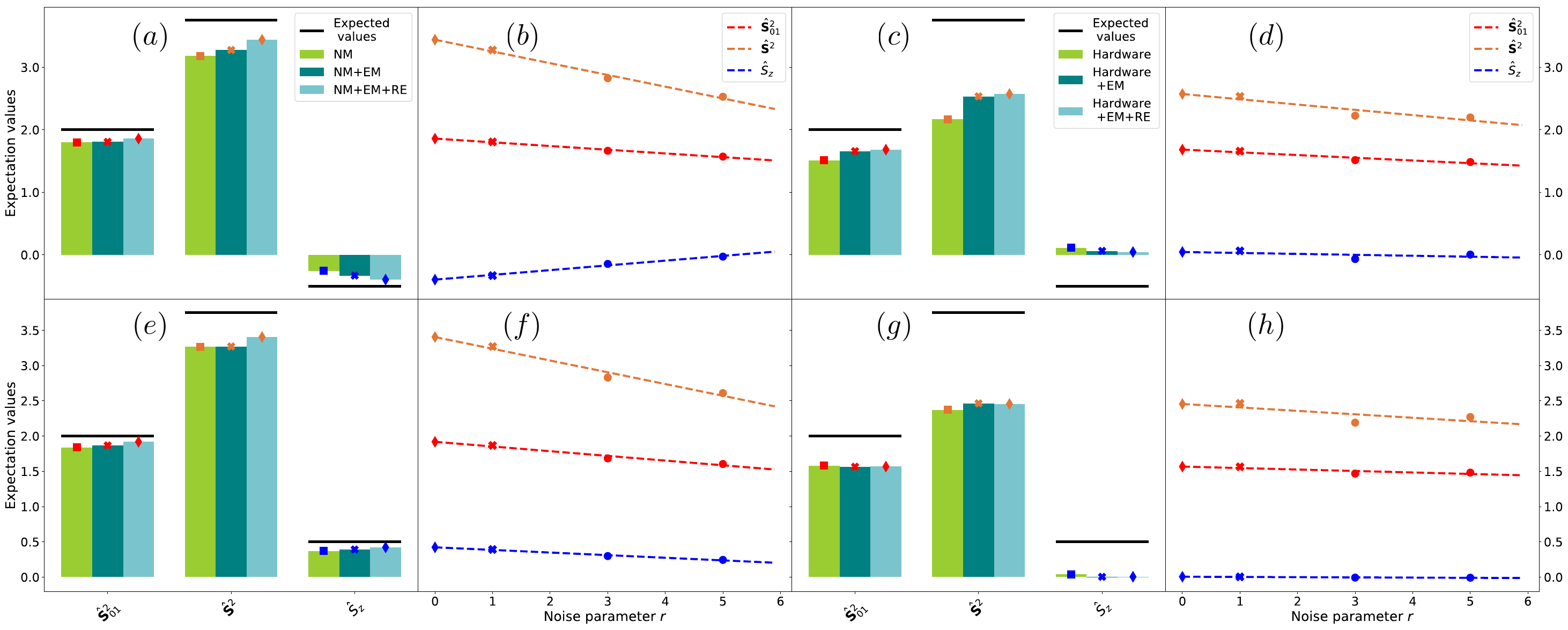}
\caption{Bar charts: expectation values of the spin operators $\hat{S}^2$, $\hat{S}_z$ and $\hat{S}_{01}^2$ for systems of $n=3$ qubits prepared in $\ket{0,1/2,-1/2}$ (14 a,c) and $\ket{1,1/2,-1/2}$ (14 e,g)
using \device{santiago} with noise model from \device{santiago} (14 a,e) and on the device \device{santiago} (14 c,g) and the time-evolution variational form.
(14 b,d,f,h): Richardson extrapolation analysis for the expectation values shown in the bar charts.The vertical scales are matched to the scales in the accompanying bar chart.}

\label{fig:FIGURE_14}
\end{figure*}

\section{Conclusions}
\label{sec:conclusions}

In this work, quantum circuits for the exact and approximate preparation of total spin eigenfunctions on quantum computers were presented.
We described two families of circuits, representative of two different approaches to address this problem.
First, we presented a recursive construction of total spin eigenfunctions based on the addition theorem of angular momentum, which 
we demonstrated for systems of 3- and 5-spin systems on graphs with triangle connectivity on IBM quantum devices.
The approach is simple to understand and implement, and it guarantees exact mapping of total spin eigenfunctions on multi-qubit wavefunctions, without making use of ancillary qubits.
On the other hand, its computational cost is at worst exponential in the number of qubits, which limits applications to large systems.

In addition, we presented a heuristic approximation of total spin eigenfunctions based on the variational optimization of a suitable cost function.
The approach does not guarantee exact mapping of total spin eigenfunctions on multi-qubit wavefunctions. On the other hand, the heuristic variational approximation lends itself to simulations on contemporary quantum hardware, as it relies on families of quantum circuits whose computational cost can by construction be fit within the computational budget allowed by a certain device.

{We show the effect of errors on the simulation of 
total spin eigenfunctions, by performing experiments on actual quantum devices.
We observed that qubit decoherence and gate errors cause significant infidelities between target and simulated wavefunctions. The extent of such a phenomenon depends on the number of spins, as well as on the simulated circuit. We note particular problems with variational simulation of the most strongly correlated states, such as the Dicke states for $n=3$ spins, as well as topologies such as $n=5$ spins in a bow-tie, with the 4-way coordinated center spin. 

Using the exact recursive construction, for $n=3$ qubits and $n=5$ qubits 
(as documented in Fig. \ref{fig:FIGURE_8} and \ref{fig:FIGURE_10}),
infidelities are $\sim$ 0.05 and $\sim$ 0.15 respectively,
and roughly consistent across various spin eigenfunctions.

Larger and less uniform infidelities are seen for $n=3$
qubits when VQE is used (Fig. \ref{fig:FIGURE_13}). These  infidelities translate into significant deviations between exact and simulated observables, as seen in Fig.~\ref{fig:FIGURE_14}.

We demonstrate the effect of state-of-the-art
error mitigation techniques to reduce the impact of measurement and gate errors on such results. To start with, systems with $3$ quits will have less error than $5$, due to the different size and topology of these systems.

Of particular significance is the performance of the Richardson extrapolation, which breaks down for some of the $n=5$ qubits using exact recursive construction (see Fig.~\ref{fig:FIGURE_9}) and some of the $n=3$ qubits with very deep variational circuits (see Fig.~\ref{fig:FIGURE_14}). In the case of $n=5$ qubits, we ascribe this effect mainly to the problem of frequency crosstalk for the central qubit of the bowtie topology, which needs to couple to $4$ other closely valued but different frequency qubits. This problem will have to be solved if we are ever going to look at strongly coupled spin qubits in spin liquids. Ideally the next system to study would be the bow-tie (Figure \ref{fig:FIGURE_6}) with the four corners (pairs (0,3) and (1,4)) also connected, with minimal connectivity 3.
}

\section*{Acknowledgments}

AC and DEG acknowledge the Università degli Studi di Milano INDACO Platform for providing resources that have contributed to the results reported in this work.
This material is based upon work supported by the U.S. Department of Energy, Office of Science, National Quantum Information Science Research Centers, Quantum Science Center.

\appendix

\section{Structure of total spin eigenfunctions}

In this Section, we list the total spin eigenfunctions explored in the present study,
for $n=3$ and $n=5$ spins, respectively in Tables \ref{tab:TABLE_4} and \ref{tab:TABLE_5}.

\begin{table}[h!]
\begin{tabular}{ccrcc}
\hline\hline
$\ell_{01}$ & $\ell$ & $\,m\,$ & & $| \ell_{01}, \ell, m \rangle$ \\
\hline
0 & $1/2$ & $-1/2$ & & $\frac{1}{\sqrt{2}} \Bigl( \ket{011}  - \ket{101} \Bigr)$ \\
0 & $1/2$ &  $1/2$ & & $\frac{1}{\sqrt{2}} \Bigl( \ket{100}  - \ket{010} \Bigr)$ \\
1 & $1/2$ & $-1/2$ & & $\frac{1}{\sqrt{6}} \Bigl( \ket{011}  + \ket{101} \Bigr) - \frac{2}{\sqrt{3}} \ket{110}$ \\
1 & $1/2$ &  $1/2$ & & $\frac{1}{\sqrt{6}} \Bigl( \ket{100}  + \ket{010} \Bigr) - \frac{2}{\sqrt{3}} \ket{001}$ \\
1 & $3/2$ & $-3/2$ & & $\ket{111}$ \\
1 & $3/2$ & $-1/2$ & & $\frac{1}{\sqrt{3}} \Bigl( \ket{011} + \ket{101} + \ket{110} \Bigr)$ \\
1 & $3/2$ &  $1/2$ & & $\frac{1}{\sqrt{3}} \Bigl( \ket{100} + \ket{010} + \ket{001} \Bigr)$ \\
1 & $3/2$ &  $3/2$ & & $\ket{000}$ \\
\hline\hline
\end{tabular}
\caption{Total spin eigenfunctions for systems of $n=3$ spins.}
\label{tab:TABLE_4}
\end{table}

\begin{table}[h!]
\begin{tabular}{ccccrcc}
\hline\hline
$\ell_L$ & $\ell_{LC}$ & $\ell_R$ & $\ell$ & & $\,m\,$ & $| \ell_L, \ell_{LC}, \ell_R, \ell, m \rangle$ \\
\hline
0 & $1/2$ & 0 & $1/2$ & & $-1/2$ & $\frac{1}{2} \Bigl(\ket{0 \, 1} - \ket{1 \, 0} \Bigr) \ket{1} \Bigl(\ket{0 \, 1} - \ket{1 \, 0} \Bigr)$ \\
0 & $1/2$ & 0 & $1/2$ & & $1/2$ & $\frac{1}{2} \Bigl(\ket{0 \, 1} - \ket{1 \, 0} \Bigr) \ket{0} \Bigl(\ket{0 \, 1} - \ket{1 \, 0} \Bigr)$ \\
0 & $1/2$ & $1$ & $1/2$ & & $-1/2$ & $\sqrt{\frac{1}{3}} \Bigl(\ket{0 \, 1} - \ket{1 \, 0} \Bigr) \Bigl[\frac{1}{2} \ket{1} \Bigl( \ket{0 \, 1} + \ket{1 \, 0} \Bigr)  - \ket{0 \, 1 \, 1} \Bigr] $ \\
0 & $1/2$ & $1$ & $1/2$ & & $1/2$ & $\sqrt{\frac{1}{3}} \Bigl(\ket{0 \, 1} - \ket{1 \, 0} \Bigr) \Bigl[\ket{1 \, 0 \, 0}-  \frac{1}{2} \ket{0} \Bigl( \ket{0 \, 1} + \ket{1 \, 0} \Bigr)  \Bigr] $ \\
$1$ & $1/2$ & 0 & $1/2$ & & $-1/2$ &  $\sqrt{\frac{1}{3}} \Bigl[\frac{1}{2} \Bigl( \ket{0 \, 1} + \ket{1 \, 0} \Bigr) \ket{1}  - \ket{1 \, 1 \, 0} \Bigr] \Bigl(\ket{0 \, 1} - \ket{1 \, 0} \Bigr)$ \\
$1$ & $1/2$ & 0 & $1/2$ & & $1/2$ & $\sqrt{\frac{1}{3}} \Bigl[\ket{0 \, 0 \, 1} -  \frac{1}{2} \Bigl( \ket{0 \, 1} + \ket{1 \, 0} \Bigr) \ket{0} \Bigr] \Bigl(\ket{0 \, 1} - \ket{1 \, 0} \Bigr)$ \\
\hline\hline
\end{tabular}
\caption{$6$ total spin eigenfunctions for systems of $n=5$ spins with $\ell_{LC}=1/2$ and $(\ell_L,\ell_R) \in \{(0,0),(0,1),(1,0)\}$.}
\label{tab:TABLE_5}
\end{table}

\section{Quantum computing terms and symbols}
\label{appendix:qis}

In Table~\ref{tab:TABLE_6}, we illustrate the quantum operations used in the present work.
Single-qubit rotations are exponentials of single-qubit Pauli operators, 
for example $R_x(\theta) = \exp(-i \theta X/2)$. 
Single-qubit Pauli operators are 
equal to special single-qubit Pauli rotations up to a global phase, for example $X = R_x(\pi/2)$.
Single-qubit operations in the Clifford group (Hadamard, $S$ and $T$ gates) are equal to special single-qubit Pauli rotations up to a global phase, namely $S = R_z(\pi/2)$, $T=R_z(\pi/4)$ 
and $H = \exp(-i \pi/2 (X+Z))$.

The $\mathsf{CNOT}$ gate is sometimes denoted $\mathsf{CNOT}_{ij}$, where $i$ and $j$ are called 
the control and target qubit respectively, and applies an $X$ transformation to its target qubit ($\oplus$ symbol) if its control qubit ($\bullet$ symbol) is in the state $|1\rangle$, the $\mathsf{cU}$ can be written as a product of up to two $\mathsf{CNOT}$ gates and four single-qubit gates, and the $\mathsf{SWAP}$ gate can be written as a product of three $\mathsf{CNOT}$ gates, $\mathsf{SWAP}_{ij} = \mathsf{CNOT}_{ij} \mathsf{CNOT}_{ji} \mathsf{CNOT}_{ij}$. 

\begin{table}[h!]
\includegraphics[width=\textwidth]{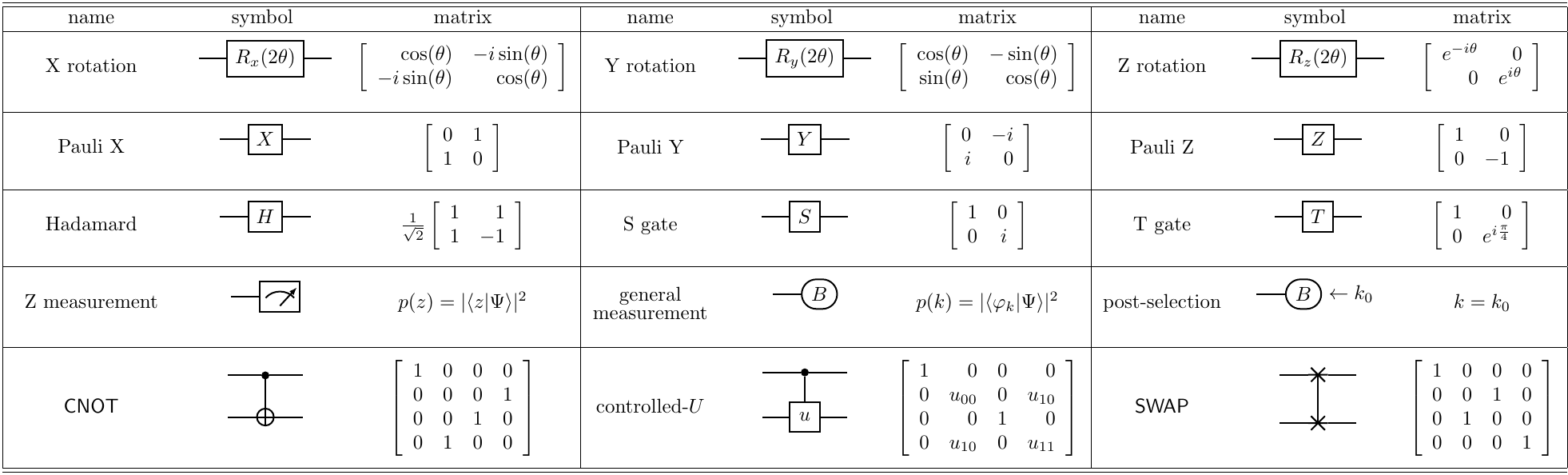}
\caption{Examples of quantum gates and circuit elements. From top to bottom: single-qubit rotations,
single-qubit Pauli operators, single-qubit operations in the Clifford group, measurements, and two-qubit gates. Adapted from Ref.~\cite{motta2021emerging}.}
\label{tab:TABLE_6}
\end{table}


%

\end{document}